# A new method for solving the Z > 137 problem and for determination of energy levels of hydrogen-like atoms

V.P.Neznamov[1], I.I.Safronov

RFNC-VNIIEF, 37 Mira Ave., Sarov, 607188 Russia

Abstract

The "catastrophe" in solving the Dirac equation for an electron in the field of a point electric charge, which emerges for the charge numbers Z > 137, is removed in this work by new method of accounting of finite dimensions of nuclei. For this purpose, in numerical solutions of equations for Dirac radial wave functions, we introduce a boundary condition at the nucleus boundary such that the components of the electron current density is zero.

As a result, for all nuclei of the periodic table the calculated energy levels practically coincide with the energy levels in standard solutions of the Dirac equation in the external field of the Coulomb potential of a point charge.

Further, for $Z > 105$, the calculated energy level functions $E(Z)$ are monotone and smooth.

The lower energy level reaches the energy $E = -mc^2$ (the electron "drop" on a nuclei) at $Z_c = 178$.

The proposed method of accounting of the finite size of nuclei can be easily used in numerical calculations of energy levels of many-electron atoms.

---

[1] E-mail: neznamov@vniief.ru

## 1. Introduction

A century ago, in 1913, Niels Bohr developed the postulates of a new quantum theory. As early as in three years, based on the theory of Bohr's orbits, A.Sommerfeld [1] developed a formula for the fine structure of energy levels of hydrogen-like atoms,

$$E = \frac{mc^2}{\left(1 + \frac{\alpha_{em}^2 Z^2}{\left(n - |\kappa| + \sqrt{\kappa^2 - \alpha_{em}^2 Z^2}\right)^2}\right)^{1/2}}. \tag{1}$$

Following the development of the Dirac theory in 1928, Dirac [2], Darwin [3] and Gordon [4] obtained expression (1) as a result of exact solution of the Dirac equation in the Coulomb field of a point charge $(-Ze)$.

In (1), $m$ is the mass of electron, $c$ is the speed of light, $\alpha_{em} = \frac{e^2}{\hbar c}$ is the electromagnetic constant of the fine structure, $Z$ is the atomic number, $n = 1, 2...$ is the main quantum number, $\kappa$ is the quantum number of the Dirac equation:

$$\kappa = \pm 1 \pm 2... = \begin{cases} -(l+1), & j = l + \frac{1}{2} \\ l, & j = l - \frac{1}{2} \end{cases}. \tag{2}$$

In (2), $j, l$ are the quantum numbers of the total and orbital momentum of the electron.

Formula (1) is became a complex number if

$$Z > \frac{|\kappa|}{\alpha_{em}} \simeq 137|\kappa|. \tag{3}$$

From the practical viewpoint of the existence of real nuclei in the periodic table, of interest in (3) are the electron states of $c|\kappa| = 1$, i.e. the $1S_{1/2}$ and $2P_{1/2}$ - states. For these states, the complexity of energy levels in (1) is often called the "Z>137 catastrophe".

It was established fairly quickly that the "catastrophe" results from the ignorance of the finite size of the nuclei.

In 1945, Pomeranchuk and Smorodinsky [5] considered an atomic system with potential

$$U = \begin{cases} -\frac{Ze^2}{r_N} & \text{для} \quad r \leq r_N \\ -\frac{Ze^2}{r} & \text{для} \quad r > r_N, \end{cases} \tag{4}$$

where $r_N$ is the nucleus radius.

As a result, they estimated $Z_c$, at which the lower energy level of the $1S_{1/2}$ - state reaches the limiting value of $E = -mc^2$.

$$Z_c = 175 \quad \text{at} \quad r_N = 0,8 \cdot 10^{-12} \text{ sm.} \tag{5}$$

This lead to an important conclusion that in the range of $Z_c \geq Z > 137$ there must exist a real function of $E(Z)$, and the "catastrophe" in (1) indeed occurs as a result of the ignorance of the finite size of nuclei.

In 1959, Zeldovich [6] demonstrated that variations in the Coulomb potential near the origin of coordinates produce minor effects on the energy spectrum of hydrogen-like atoms.

An overview of subsequent papers devoted to the structure of hydrogen-like atoms at $Z\alpha > 1$ is presented in the papers [7], [9], [10].

In [7], to analyze the structure of energy levels, in addition to the potential (4) Zeldovich and Popov used a potential corresponding to the potential of a uniformly charged sphere.

$$U_1 = \begin{cases} \dfrac{Ze^2}{r_N}\left(-\dfrac{3}{2} + \dfrac{1}{2}\left(\dfrac{r}{r_N}\right)^2\right) & \text{for } r \leq r_N \\ -\dfrac{Ze^2}{r} & \text{for } r > r_N \end{cases} \tag{6}$$

The authors [8] numerically calculated the energy levels of the first nine states $\left(1S_{1/2}, 2S_{1/2}......3D_{5/2}\right)$ as a function of $Z$ for the potential (6). The value of $Z_c$ determined in [8] is $Z_c = 169$ for $r_N = 9,5 \cdot 10^{-12}$ cm. This value is close to the values of $Z_c = 170 \div 175$ obtained by other researchers (see [5], [7]).

Now it is known more than 30 electrostatic potentials, which take into account the finite distribution of electric charge in the atom nuclei, have been offered by the different authors. These potentials are used in the various machine codes for determination of electronic structure of atoms and molecules. A review of developed potentials and their use in numerical calculations of Dirac and Schrödinger equations are in [9] (see also [10]). In the reviews [9], [10] there are also the wide range of literature of analytical and numerical determination of electronic structure of atoms and molecules.

The solutions of the Dirac and Schrödinger equations with using of the finite potentials of the atomic nucleus are determined by the standard method. Firstly the wave function of electrons are calculated within a nucleus in a filed of electrostatic potential of interest. Then, the values of these functions at the boundary of the nucleus are equated with the similar values of wave function of electrons in the Coulomb field. The boundary condition for radial wave at functions $r \to \infty$ and $r = 0$ determine the energy spectrum of the atomic and molecular systems.

According to [5] - [9] the use of different introduced electrostatic potentials of atomic nuclei lead to relatively little change (tenths of a percent) both absolute values of the energy levels, and differences of the energy levels. These changes grow when $Z$ increases.

In the present paper, the problem of determination of the energy spectrum of hydrogen-like atoms, including nucleus with $Z > 137$, is solved by a new approach to numerical calculations of the Dirac equation in the Coulomb field by introducing a boundary condition for wave functions at the boundary of the nuclei of interest.

The boundary condition at the nucleus boundary is taken by analogy with the analysis of the possibility of existence of stationary bound states in the Schwarzschild gravitational field [11]. It involves zeroing of the $\varphi$-component of Dirac current density at the boundary of the nucleus of interest, which resolves itself in zeroing of one of two radial wave functions at the nucleus boundary in the Coulomb field. In this case the calculations are simplified and its are made from $r \to \infty$ up to the boundary of the nucleus $r_N$.

This paper has the following structure. For completeness of presentation, Section 2 contains the Dirac equation in the Coulomb field, outlines the procedure of separation of variables, and gives a system of equations for radial wave functions.

Section 3 explores the behavior of the components of the vector of current density of Dirac particles and introduces the boundary condition for wave functions at the boundary of the nucleus.

Section 4 reviews the results of numerical calculations of energy spectra of hydrogen-like atoms with various $Z$.

The Conclusion summarizes the results of this study.

## 2. Dirac equation in the Coulomb field of the charge $(-Ze)$

Below we will use the system of units $\hbar = c = 1$, the signature

$$g^{\alpha\beta} = diag[1,-1,-1,-1]; \qquad (7)$$

$\beta, \alpha^k$, $k = 1,2,3$ are 4x4 Dirac matrices in the Dirac-Pauli representation, and $\sigma^k$ are 2x2 Pauli matrices.

We consider the stationary case, when the wave function can be written as

$\psi(\mathbf{r},t) = \psi(\mathbf{r})e^{-iEt}$.

The Dirac equation in the Coulomb field of the point charge $(-Ze)$ in spherical coordinates $(r,\theta,\varphi)$ can be expressed as:

$$E\psi(\mathbf{r}) = \left[\beta m - i\alpha^1\left(\frac{\partial}{\partial r} + \frac{1}{r}\right) - i\alpha^2 \frac{1}{r}\left(\frac{\partial}{\partial \theta} + \frac{1}{2}\operatorname{ctg}\theta\right) - i\alpha^3 \frac{1}{r\sin\theta}\frac{\partial}{\partial \varphi} - \frac{Ze^2}{r}\right]\psi(\mathbf{r}).$$

(8)

Eq. (8) allows for the separation of variables, if the bispinor $\psi(\mathbf{r}) = \psi(r,\theta,\varphi)$ is given by

$$\psi(r,\theta,\varphi) = \begin{pmatrix} F(r)\xi(\theta) \\ -iG(r)\sigma^3\xi(\theta) \end{pmatrix} e^{im_\varphi \varphi}$$

(9)

and the following equation is used (see, e.g., [12]):

$$\left[-\sigma^2\left(\frac{\partial}{\partial\theta} + \frac{1}{2}\operatorname{ctg}\theta\right) + i\sigma^1 m_\varphi \frac{1}{\sin\theta}\right]\xi(\theta) = i\kappa\xi(\theta).$$

(10)

In (9), (10), $\xi(\theta)$ are spherical harmonics for spin ½, $m_\varphi$ is the magnetic quantum number, $\kappa$ is the quantum number (2).

$\xi(\theta)$ can be represented as in [13].

$$\xi(\theta) = \begin{pmatrix} Y_{jm_\varphi}^{-\frac{1}{2}}(\theta) \\ Y_{jm_\varphi}^{\frac{1}{2}}(\theta) \end{pmatrix} = (-1)^{m_\varphi + \frac{1}{2}} \sqrt{\frac{1}{4\pi}\frac{(j-m_\varphi)!}{(j+m_\varphi)!}} \begin{pmatrix} \cos\frac{\theta}{2} & \sin\frac{\theta}{2} \\ -\sin\frac{\theta}{2} & \cos\frac{\theta}{2} \end{pmatrix} \times$$

$$\times \begin{pmatrix} \left(\kappa - m_\varphi + \frac{1}{2}\right) P_l^{m_\varphi - \frac{1}{2}}(\theta) \\ P_l^{m_\varphi + \frac{1}{2}}(\theta) \end{pmatrix}.$$

(11)

In (11), $P_l^{m_\varphi \pm \frac{1}{2}}(\theta)$ are Legendre polynomials.

The separation of variables gives a system of equations for real radial functions $F(r), G(r)$. We write these equations in dimensionless variables $\varepsilon = \frac{E}{m}$, $\rho = \frac{r}{l_c}$, where $l_c = \frac{\hbar}{mc}$ is the Compton wavelength of the electron.

$$\frac{dF}{d\rho} + \frac{1+\kappa}{\rho}F - \left(\varepsilon + 1 + \frac{\alpha_{em} Z}{\rho}\right)G = 0$$

$$\frac{dG}{d\rho} + \frac{1-\kappa}{\rho}G + \left(\varepsilon - 1 + \frac{\alpha_{em} Z}{\rho}\right)F = 0.$$

(12)

If we introduce the phase from the definition

$$\operatorname{tg}\Phi = \frac{F(\rho)}{G(\rho)},$$

(13)

then the energy spectrum $\varepsilon_n$ can also be defined from the equation for the phase $\Phi = \text{arctg}\dfrac{F(\rho)}{G(\rho)} + k\pi, \ k = 0, \pm 1, \pm 2,...$ in the form proposed by Vronsky [11]

$$\frac{d\Phi}{d\rho} = \varepsilon + \frac{\alpha_{em}Z}{\rho} + \cos 2\Phi - \frac{\kappa}{\rho}\sin 2\Phi. \tag{14}$$

For the finite motion of the electron, asymptotics of solutions to Eqs. (12) for $\rho \to \infty$ is given by

$$F(\rho) = C_1 e^{-\rho\sqrt{1-\varepsilon^2}}$$
$$G(\rho) = -\sqrt{\frac{1-\varepsilon}{1+\varepsilon}}F(\rho). \tag{15}$$

The phase $\Phi$ for $\rho \to \infty$ equals

$$\Phi = -\text{arctg}\sqrt{\frac{1+\varepsilon}{1-\varepsilon}}. \tag{16}$$

### 3. Electron current density, boundary condition for the wave functions

In the course of separation of variables when deriving Eqs. (10), (12) from Eq. (8), we performed an equivalent substitution of the Dirac matrices

$$\alpha^1 \to \alpha^3; \ \alpha^2 \to \alpha^1; \ \alpha^3 \to \alpha^2. \tag{17}$$

Then, considering (9), (11), components of the Dirac current density equal

$$j^r = \psi^+\alpha^3\psi = -iF(\rho)G(\rho)\left[\xi^+(\theta)(\sigma^3\sigma^3 - \sigma^3\sigma^3)\xi(\theta)\right] = 0, \tag{18}$$

$$j^\theta = \psi^+\alpha^1\psi = -2F(\rho)G(\rho)\left[\xi^+(\theta)\sigma^2\xi(\theta)\right] = 0, \tag{19}$$

$$j^\varphi = \psi^+\alpha^2\psi = 2F(\rho)G(\rho)\left[\xi^+(\theta)\sigma^1\xi(\theta)\right] \neq 0. \tag{20}$$

The equalities (18) – (20) coincide with previously obtained results in [14].

Our boundary condition involves zeroing of the current component $j^\varphi$ at the nucleus boundary $\rho_N$, which resolves itself into zeroing of one of the two wave functions $F(\rho_N)$, $G(\rho_N)$:

$$F(\rho_N)G(\rho_N) = 0. \tag{21}$$

The boundary condition (21) is similar to the condition near the "event horizon" introduced in the numerical calculations of the solution to the Dirac equation in the Schwarzschild field [11].

As a result, for the values of the gravitational coupling constant $\alpha \ll 1$, calculations [11] yield energy levels close to the energy levels in the hydrogen atom.

# 4. Results of numerical calculations of the energy spectrum of hydrogen-like atoms with effective accounting of the finite size of nuclei

In the calculations, the size of nuclei were determined from the relationships

$$r_N = \left(0,836 A^{1/3} + 0,57\right) \cdot 10^{-13} \text{ см}, \quad A > 9 \quad [16] \tag{22}$$

$$r_N = 1,3 \cdot 10^{-13} \cdot A^{1/3} \text{ см}, \quad A < 9$$

In (22), $A$ is the atomic weight of the nucleus.

The equation for phase (14) was solved by the fifth-order Runge-Kutta implicit method with step control [15]. We used the Ila scheme to obtain the three-stage Rado IIA method.

From two possible variants of implementation of condition (21), we will fulfil it, like in [11], using equality

$$G(\rho_N) = 0. \tag{23}$$

Some reason for this is known smallness of function $G(\rho_N)$ in comparison with function $F(\rho)$ in nonrelativistic approximation of Dirac equation.

It follows from (23) that the condition for the phase equals

$$\Phi(\varepsilon, \kappa, Z) = k\frac{\pi}{2}, \quad k = \pm 1, \pm 3, \pm 5... \tag{24}$$

Tables 1 – 3 contain the values of energy levels for the hydrogen atom $Z = 1, \ A = 1$ obtained by numerical calculations of Eq. (14) with the boundary conditions (16), (24) for $\kappa = \pm 1, \pm 2, \pm 3$ and $n = 1 \div 11$.

The tables also present corresponding energy values obtained from (1) and relative deviations of calculated values from analytical ones in percent.

Table 1. Energy levels of the hydrogen atom for the $S_{1/2}$, $P_{1/2}$ - states $(\kappa = \pm 1)$.

| n | $1-\varepsilon_{an}$ | $1-\varepsilon_{num}$ | $\delta(\%)$ | Comment |
|---|---|---|---|---|
| 1 | 2.6640E-05 | 2.6641E-05 | -0.004 | No solution available for $\kappa = +1$ |
| 2 | 6.6600E-06 | 6.6602E-06 | -0.003 | |
| 3 | 2.9600E-06 | 2.9601E-06 | -0.003 | |
| 4 | 1.6650E-06 | 1.6651E-06 | -0.006 | |
| 5 | 1.0656E-06 | 1.0656E-06 | 0.000 | |
| 6 | 7.4000E-07 | 7.3999E-07 | 0.001 | |
| 7 | 5.4367E-07 | 5.4367E-07 | 0.000 | |
| 8 | 4.1625E-07 | 4.1624E-07 | 0.002 | |
| 9 | 3.2889E-07 | 3.2888E-07 | 0.002 | |
| 10 | 2.6640E-07 | 2.6639E-07 | 0.003 | |
| 11 | 2.2016E-07 | 2.2015E-07 | 0.006 | |

Table 2. Energy levels of the hydrogen atom for the $P_{3/2}$, $D_{3/2}$ - states $(\kappa = \pm 2)$.

| n | $1-\varepsilon_{an}$ | $1-\varepsilon_{num}$ | $\delta(\%)$ | Comment |
|---|---|---|---|---|
| 2 | 6.6599E-06 | 6.6585E-06 | 0.022 | No solution available for $\kappa = +2$ |
| 3 | 2.9600E-06 | 2.9603E-06 | -0.009 | |
| 4 | 1.6650E-06 | 1.6653E-06 | -0.016 | |
| 5 | 1.0656E-06 | 1.0656E-06 | 0.004 | |
| 6 | 7.3999E-07 | 7.3997E-07 | 0.004 | |
| 7 | 5.4367E-07 | 5.4367E-07 | 0.001 | |
| 8 | 4.1625E-07 | 4.1622E-07 | 0.007 | |
| 9 | 3.2889E-07 | 3.2887E-07 | 0.006 | |
| 10 | 2.6640E-07 | 2.6637E-07 | 0.012 | |
| 11 | 2.2016E-07 | 2.2017E-07 | -0.001 | |

Table 3. Energy levels of the hydrogen atom for the $D_{5/2}$, $F_{5/2}$ - states $(\kappa = \pm 3)$.

| n | $1-\varepsilon_{an}$ | $1-\varepsilon_{num}$ | $\delta(\%)$ | Comment |
|---|---|---|---|---|
| 3 | 2.9600E-06 | 2.9597E-06 | 0.011 | No solution available for $\kappa = +3$ |
| 4 | 1.6650E-06 | 1.6652E-06 | -0.010 | |
| 5 | 1.0656E-06 | 1.0657E-06 | -0.006 | |
| 6 | 7.3999E-07 | 7.3997E-07 | 0.004 | |
| 7 | 5.4367E-07 | 5.4367E-07 | 0.000 | |
| 8 | 4.1625E-07 | 4.1622E-07 | 0.007 | |
| 9 | 3.2889E-07 | 3.2887E-07 | 0.006 | |
| 10 | 2.6640E-07 | 2.6637E-07 | 0.012 | |
| 11 | 2.2016E-07 | 2.2017E-07 | -0.001 | |

We can see that the calculated and analytical values of energy values are in close agreement to within hundredths of percent $\left(\delta = \dfrac{\varepsilon_{num.} - \varepsilon_{an.}}{\varepsilon_{an.}} \lesssim 10^{-4}\right)$.

Within the above accuracy, the calculations reproduce degeneration of the energy levels with the same total momentum $j$ (the same value of $|\kappa|$) typical for the fine-structure formula (1).

Next, energy levels of the one-electron atoms were calculated for the following nuclei: $B\,(Z=5, A=10)$, $Ne\,(Z=10, A=21)$, $Mn\,(Z=25, A=5)$, $Sn\,(Z=50, A=119)$, $U\,(Z=92, A=238)$, $(Z=104, A=261)$. For hypothesized nuclei, $Z > 104$, the ratio $\dfrac{A}{Z}$ was chosen equal to $2,9$.

The results of the calculations for three lower levels and for the values of $\kappa = \pm 1, \pm 2, \pm 3$ are shown in Figs. 1 – 6. For comparison, the same figures present some numerical results [8] and analytical values from the fine-structure formula (1). In the calculations [8] the nucleus radiuses were determined from the relationship $r_N = 1,2 \cdot 10^{-13} \cdot A^{1/3}$ cm.

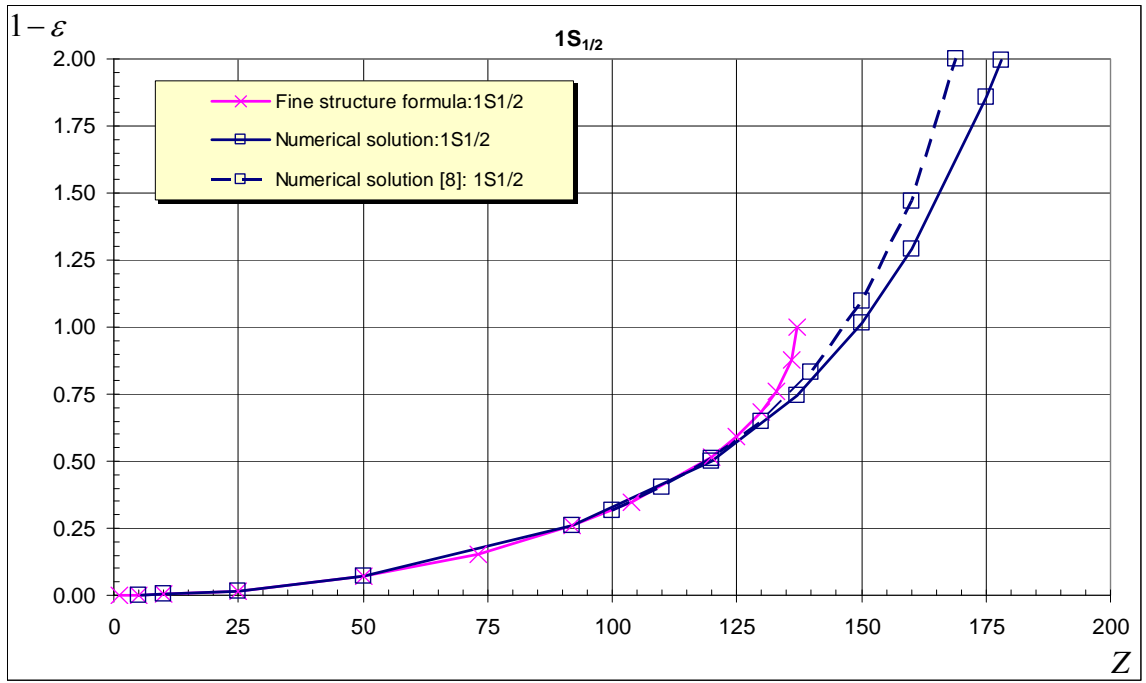

Fig. 1. The plots of $E(Z)$ for the $1S_{1/2}$ - state.

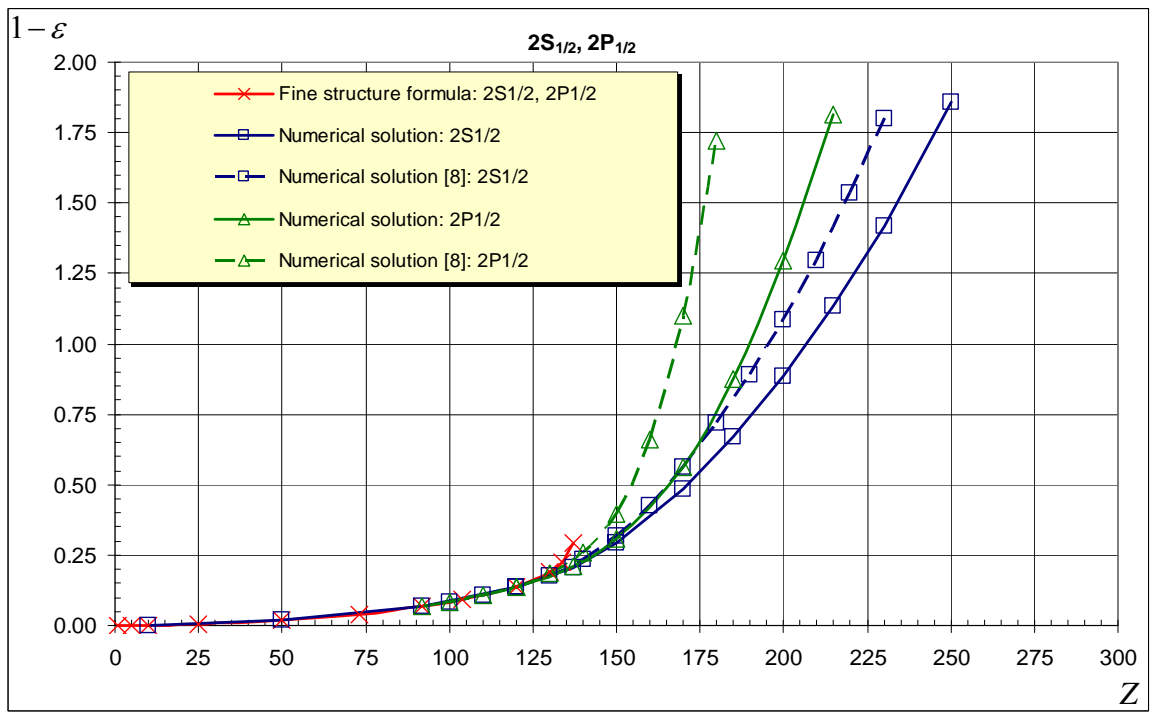

Fig. 2. The plots of $E(Z)$ for the $2S_{1/2}$, $2P_{1/2}$ - states.

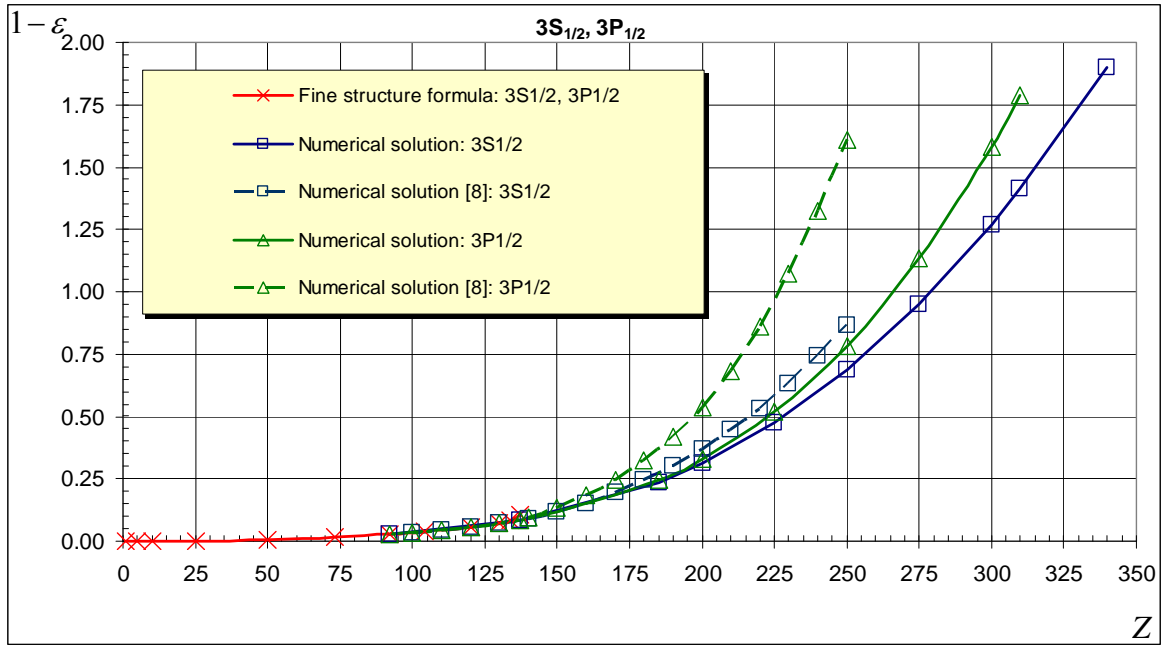

Fig. 3. The plots of $E(Z)$ for the $3S_{1/2}$, $3P_{1/2}$ - states.

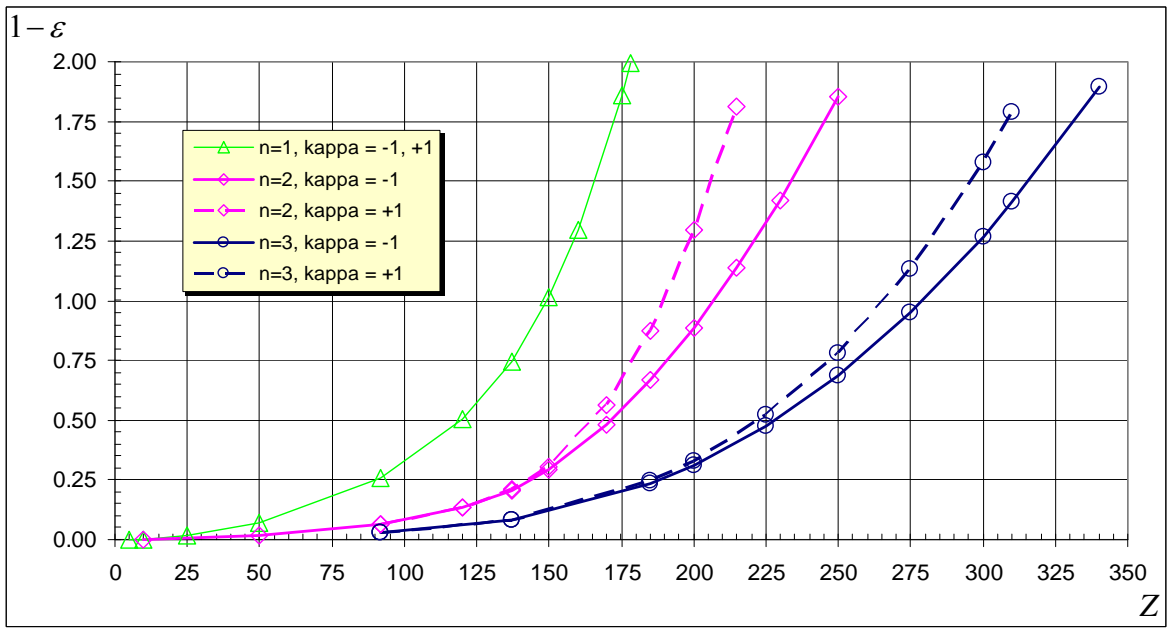

Fig. 4. Calculated plots of $E(Z)$ for the states with $n = 1, 2, 3$ and $\kappa = \pm 1$

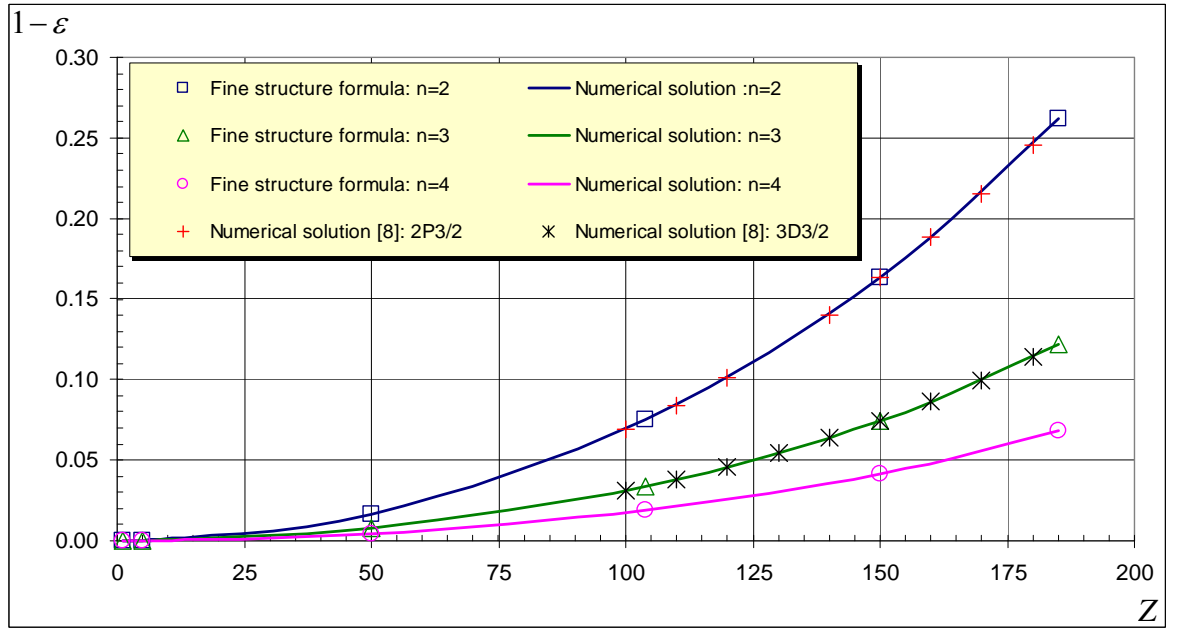

Fig. 5. The plots of $E(Z)$ for $P_{3/2}$, $D_{3/2}$ - states and $n = 2, 3, 4$

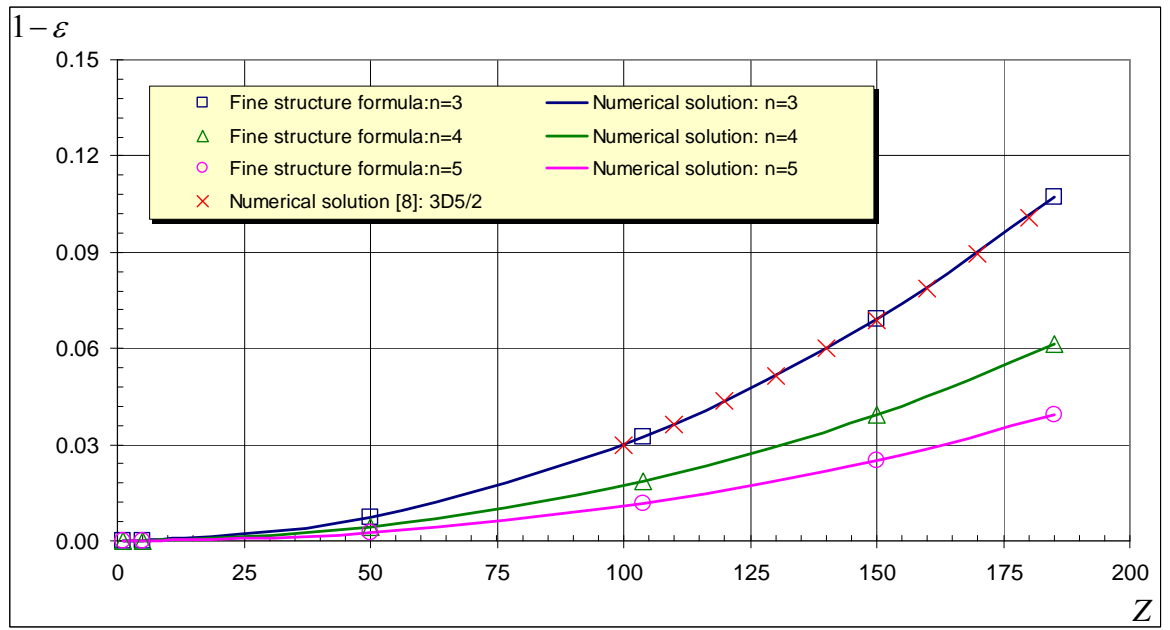

Fig. 6. The plots of $E(Z)$ for the $D_{5/2}$, $F_{5/2}$ - states and $n = 3, 4, 5$.

These results indicate that the formula (1) is in a good agreement with the calculated values of energy levels for all the known elements of the periodic table.

For $\kappa = -1$ $\left(1S_{1/2}\right)$, any noticeable discrepancy for the lower level $(>1\%)$ occurs at $Z > 105$ (Fig. 1).

The calculated plots of $E(Z)$ are smooth and monotone.

The lower level $1S_{1/2}$ reaches the value of $\varepsilon = -1$ (the electron "drop" on a nuclei) at $Z_c \simeq 178$.

If the level $1S_{1/2}$ reaches the lower continuum $\varepsilon = -1$ at $Z > 178$, one must move from single-body quantum mechanics to many-body quantum field theory [7].

In this paper, the plots of $E(Z)$ for $Z > 178$ are shown in Figs. 2 – 4 for methodological reasons. These plots have no singularities and are qualitatively similar to the plots of $E(Z)$ for the lower energy level $1S_{1/2}$.

In accordance with the results obtained in [7], [8] in Figs. 2 - 4 one can see that energy levels with the same $j$ are no more degenerate for $Z > 137$.

As the values of $n$ and $\kappa$ grow, the values of $Z$, at which energy levels with the same $j$ begin to differ, get higher. It follows from Figs. 5, 6 that the levels $P_{3/2}$, $D_{3/2}$ and $D_{5/2}$, $F_{5/2}$ coincide up to $Z_c = 178$. For these levels, one can also see good agreement with the fine-structure formula.

As a result of effective accounting of the finite size of nuclei using the boundary condition for the Dirac wave functions (21), (24), energy levels for $Z \leq 105$ practically coincide with the fine-structure formula (1) and with the results in [7], [8] using effective nucleus potentials (4), (6).

It means an absence of appreciable effect of a values of electron location probability in a nucleus on the energy spectrum (maximum probability - in the calculations with the use of the singular Coulomb potential; smaller probability – in the calculations with the use of the finite electrostatic potentials of nuclei; zero probability – in the calculation of this paper with the use of the boundary condition (21)).

For $Z > 105$, the plots of $E(Z)$ based on the results of this work are less steep (see Figs. 1 - 6). This leads to a somewhat higher value of $Z_c = 178$ compared to the values of $Z_c \approx 170$ in [7], [8]. The difference between the plots of $E(Z)$ decreases as the quantum numbers $n$ and $\kappa$ grow.

A single-body quantum-mechanical consideration becomes more approximate when $Z$ increases. It is necessary to take into account the effects of quantum electrodynamics and using of a many-body relativistic quantum theory of heavy and superheavy nuclei. Considering this fact, a value $Z_c = 178$ derived in this paper with the boundary condition (21), which provides the zero probability of electron location within a nucleus, one should consider as the upper limit of a true value $Z_c$. In [10] there is a formulation of conditions which must be fulfilled for determination $Z_c$ in future experiments.

## 5. Conclusions

The calculations to determine energy levels of hydrogen-like atoms with effective accounting (21) of the finite size of nuclei allow us to draw the following conclusions:

1. Calculations with $Z=1, A=1$ reproduce the fine-structure formula (1) for the hydrogen atom to within $\sim 10^{-4}$.
2. The calculations are in a good agreement with the fine-structure formula for all the known nuclei of the periodic table. For the lower level, any noticeable discrepancy occurs at $Z > 105$.
3. The calculated plots of $E(Z)$ are smooth and monotone.
4. The lower level $1S_{1/2}$ reaches the value of $\varepsilon = -1$ ($E = -mc^2$ is the electron "drop" on a nuclei) at $Z_c = 178$.
5. To account of the finite size of nuclei, the boundary condition (21), which shows well for the one-electron case, can be easily applied to calculations of many-electron atoms using solutions of the Dirac equation for radial wave functions.

### Acknowledgement

We thank our colleagues professors P.P. Fiziev, M.A. Vronsky and A.A. Sadovy for fruitful discussions, and A.L. Novoselova for her significant technical help in the preparation of the manuscript.